\documentclass{jpp}
\usepackage{graphicx}
\usepackage{epstopdf, epsfig}
\usepackage{amsmath}	% Advanced maths commands
\usepackage{amssymb}	% Extra maths symbols
\usepackage{multicol}        % Multi-column entries in tables
\usepackage{pdflscape}	% Landscape pages
\usepackage{gensymb}
\usepackage{tikz}
\usetikzlibrary{arrows}

\renewcommand{\eqref}[1]{Eq.~(\ref{#1})}

\newcommand{\bea}{\begin{eqnarray}}
\newcommand{\eea}{\end{eqnarray}}
\newcommand{\beq}{\begin{equation}}
\newcommand{\eeq}{\end{equation}}

\newcommand{\tA}{\tau_{\rm A}}
\newcommand{\vA}{v_{\rm A}}

\newcommand{\tnl}{\tau_{\rm nl}}

\newcommand{\vB}{\mathbf{B}}

\newcommand{\vu}{\mathbf{u}}
\newcommand{\vb}{\mathbf{b}}

\newcommand{\vz}{\mathbf{z}}

\newcommand{\vzp}{\vz_\perp}

\newcommand{\lpar}{l_\parallel}
\newcommand{\Lpar}{L_\parallel}

\newcommand{\hlpar}{\hat{l}_\parallel}
\newcommand{\hlam}{\hat{\lambda}}

\newcommand{\SL}{S_{L_\perp}}

\newcommand{\hld}{\hlam_{\rm D}}
\newcommand{\lamd}{\lambda_{\rm D}}

\newcommand{\hldpm}{\hat{\lambda}_{{\rm D},i-1}}
\newcommand{\hldp}{\hat{\lambda}_{{\rm D},i}}

\newcommand{\zom}{{\zeta_n^\perp/n}}

\newcommand{\qd}{{q_{\rm D}}}
\newcommand{\tc}{\tau_{\rm C}}

\shorttitle{Disruption of Alfv\'enic turbulence by reconnection in collisionless plasma}
\shortauthor{A. Mallet, A.~A. Schekochihin and B. D. G. Chandran}

\title{Disruption of Alfv\'enic turbulence by magnetic reconnection in a collisionless plasma}

\author{Alfred Mallet\aff{1}
  \corresp{\email{alfred.mallet@unh.edu}},
  A. A. Schekochihin\aff{2,3}
 \and B. D. G. Chandran\aff{1}}

\affiliation{\aff{1}Space Science Center, University of New Hampshire, Durham, NH 03824, USA
\aff{2}Rudolf Peierls Centre for Theoretical Physics, University of Oxford, Oxford OX1 3NP, UK
\aff{3}Merton College, Oxford OX1 4JD, UK}

\begin{document}

\maketitle

\begin{abstract}
We calculate the disruption scale $\lamd$ at which sheet-like structures in dynamically aligned Alfv\'enic turbulence are destroyed by the onset of magnetic reconnection in a low-$\beta$ collisionless plasma. 
The scaling of $\lamd$ depends on the order of the statistics being considered, with more intense structures being disrupted at larger scales. 
The disruption scale for the structures that dominate the energy spectrum is $\lamd\sim L_\perp^{1/9}(d_e\rho_s)^{4/9}$, where $d_e$ is the electron inertial scale, $\rho_s$ is the ion sound scale, and $L_\perp$ is the outer scale of the turbulence. 
When $\beta_e$ and $\rho_s/L_\perp$ are sufficiently small, the scale $\lamd$ is larger than $\rho_s$ and there is a break in the energy spectrum at $\lamd$, rather than at $\rho_s$. 
We propose that the fluctuations produced by the disruption are circularised flux ropes, which may have already been observed in the solar wind. We predict the relationship between the amplitude and radius of these structures and quantify the importance of the disruption process to the cascade in terms of the filling fraction of undisrupted structures and the fractional reduction of the energy contained in them at the ion sound scale $\rho_s$. Both of these fractions depend strongly on $\beta_e$, with the disrupted structures becoming more important at lower $\beta_e$.
Finally, we predict that the energy spectrum between $\lamd$ and $\rho_s$ is steeper than $k_\perp^{-3}$, when this range exists. Such a steep ``transition range" is sometimes observed in short intervals of solar-wind turbulence.
The onset of collisionless magnetic reconnection may therefore significantly affect the nature of plasma turbulence around the ion gyroscale.
\end{abstract}

\section{Introduction}
Astrophysical plasmas are often turbulent, with power-law spectra over a wide range of scales. In many situations, a strong background magnetic field $\vB_0$ can be assumed, and often the plasma is only weakly collisional. A well-studied example of such a system is the solar wind, in which the turbulence is directly measured by spacecraft \citep{bruno2013,chen2016}. The nature of the turbulence depends on how the scale of interest compares to the ion gyroradius $\rho_i =v_{{\rm th} i}/\Omega_i$, where the ion thermal speed $v_{{\rm th} i} = \sqrt{2T_i/m_i}$ and the ion gyrofrequency $\Omega_i = ZeB_0/m_i c$. Regardless of whether the plasma is collisional or collisionless, on length scales much larger than the ion gyroradius, $k_\perp\rho_i\ll1$, Alfv\'enically polarized fluctuations obey the RMHD equations \citep{kadomtsev1973,strauss1976,schektome2009}, which describe nonlinearly interacting Alfv\'en wavepackets (represented by the Elsasser fields $\vzp^\pm = \vu \pm \vb$) propagating up and down the background magnetic field at the Alfv\'en speed $\vA = B_0/\sqrt{4\pi m_i n_i}$. At smaller, ``kinetic" scales, $k_\perp\rho_i\gtrsim1$, the Alfv\'en waves become dispersive ``kinetic Alfv\'en waves" (as confirmed in the solar wind: see \citealt{chen2013}).%\footnote{In the low $\beta_e$ situation we will consider in this paper, the scale at which the Alfv\'enic fluctuations become dispersive is in fact approximately $\sqrt{\rho_i^2/2+\rho_s^2}$, where $\rho_s=\rho_i\sqrt{ZT_e/2T_i}$ is the ion sound scale \citep{zocco2011}.}. 

The structure of strong RMHD turbulence at large scales, $k_\perp\rho_i\ll1$, is relatively well understood. First, the fluctuations are ``critically balanced" \citep{gs95,gs97,rcb} -- their linear timescale $\tA \sim \lpar / \vA$ and nonlinear timecale $\tnl$ are comparable (here $\lpar$ is the parallel coherence length). This leads to anisotropic fluctuations with $\lpar \gg \lambda$, where $\lambda \sim 1/k_\perp$ is the perpendicular coherence scale. Second, at least in numerical simulations \citep{mcatbolalign,pmbc}, the fluctuations dynamically ``align" so that the vector velocity and magnetic-field perturbations point in the same direction up to a small, scale-dependent angle $\theta$ \citep{boldyrev,Chandran14,ms16}. This causes the fluctuations to become anisotropic within the perpendicular plane, with scale $\xi \gg \lambda$ in the direction of the vector-field perturbations. Together, these two phenomena mean that the turbulent structures are 3D anisotropic, with $\lpar \gg\xi\gg\lambda$. This anisotropy has been measured both in numerical simulations \citep{verdini2015,mallet3d} and in the solar wind \citep{chen3d}, and results in the turbulent structures becoming increasingly sheet-like at smaller scales. We review scalings obtained in a simple model of this type of Alfv\'enic turbulence by \cite{ms16} in Section \ref{sec:turb}. 
At smaller scales $\lambda \lesssim \rho_i$, the turbulence is also likely to be critically balanced \citep{cho2004,schektome2009,boldyrevkaw2012,tenbarge2012} and has a steeper perpendicular spectral index of approximately $-2.8$ \citep{alexandrova2009,chenkaw2010,sahraoui2010}.

Since sheet-like structures are generically unstable to the tearing mode and the onset of magnetic reconnection, the formation of such structures by the large-scale Alfv\'enic turbulence immediately suggests that at some scale, the reconnection process may become faster than the dynamically aligning cascade, and disrupt the sheet-like structures. In resistive RMHD, the disruption scale was calculated by \cite{msc_disruption} and \cite{loureiroboldyrev} as $\lambda_{\rm D} \sim L_\perp \SL^{-4/7}$, where $\SL \doteq L_\perp \overline{\delta z} / \eta$ is the outer-scale Lundquist number (equivalently, the magnetic Reynolds number), $\eta$ being the Ohmic diffusivity (resistivity)%\footnote{This scaling is obtained for the fluctuations which dominate the energy spectrum ($n=2$, in the language of Section \ref{sec:turb})-- higher-amplitude fluctuations disrupt at a larger scale, since they are more highly aligned; see \cite{msc_disruption} for details.}
. At scale $\lamd$, the sheet-like structures reconnect, and 
%this process is fast enough that the nonlinear saturated state of the instability is reached -- 
%the initial sheet-like structure is
are converted into %
%a set of 
circularised flux ropes with radius $\lamd$, destroying the dynamic alignment. Below $\lambda_{\rm D}$, \cite{msc_disruption} proposed that these flux-rope-like structures realign and are disrupted again in a recursive fashion, leading to a steeper spectrum of approximately $k_\perp^{-11/5}$ and a final dissipative cutoff scale of $\lambda_\eta \sim L_\perp\SL^{-3/4}$\footnote{\citet{bl2017} agree with these scalings of the spectrum and the dissipative cutoff but do not believe that tearing-produced islands can fully circularize.}. This quantifies the role that reconnection plays in the dynamics of MHD turbulence, a topic that has a long history \citep{matthaeuslamkin86,politano89,retino2007,sundkvist2007,servidio2009,zhdankin13,osman14rec,greco2016,cerri2017,cerri2017b,franci2017}.

Here, we extend the \cite{msc_disruption} model of the disruption of Alfv\'enic turbulence by reconnection to the low-$\beta_e$ collisionless case, where the reconnection is enabled by electron inertia, rather than resistivity. The nature of the tearing mode in this regime is reviewed in Section \ref{sec:tear}. %, and the nature of the tearing mode is different. 
Our main conclusion, arrived at in Section \ref{sec:scale}, is that for sufficiently low electron beta $\beta_e=8\pi n_e T_e / B_0^2$ and large enough separation between the ion sound scale $\rho_s = \rho_i\sqrt{ZT_e/2T_i}\sim\rho_i$ \textbf{(we are assuming that $T_e \sim T_i$)} and the outer scale $L_\perp$, the onset of reconnection may cause the turbulence to be disrupted, inducing a spectral break at a scale $\lamd$ larger than the scale $\sqrt{\rho_i^2/2+\rho_s^2} \sim \rho_i\sim \rho_s$ at which the Alfv\'en waves in this regime become dispersive \citep{zocco2011}. 
This means that the turbulent structures around the ion scale, which are the starting point for the kinetic-Alfv\'en-wave turbulent cascade at smaller scales, are created by tearing-induced disruption of the large-scale sheets produced by the RMHD turbulent dynamics, rather than solely by the change in the dispersion relation governing the linear wave response \citep[cf.][]{cerri2017,franci2017}.

In the solar wind at 1AU, where $\beta_e \sim 1$, we predict that only the most intense sheet-like structures are disrupted and converted into flux ropes. Interestingly, ``Alfv\'en vortices", which appear to be very similar to the flux-rope structures, have already been observed even in the solar wind at 1AU \citep{lion2016,perrone2016}. The mechanism proposed in this paper is a physical way to generate these structures. In Section~\ref{sec:stat}, we derive the fractional reduction in the volume filled by and energy contained within undisrupted, sheet-like structures at the ion sound scale as a function of $\beta_e$, showing that both these fractions decrease as $\beta_e$ decreases. We also derive the dependence of the amplitude of the newly formed flux ropes on their scale -- this could be compared with the observed Alfv\'en vortices. Closer to the Sun, in the region to be explored by the Parker Solar Probe \citep{spp2016}, it is expected that, at least in fast-solar-wind streams, $\beta_e \approx 0.01$ \citep{chandran2011}, in which case the moderate-amplitude structures that dominate the energy spectrum may be disrupted. Thus, our results may be especially relevant to the turbulence that will be observed by this new mission. In Section~\ref{sec:disrange}, we derive approximate scalings for the energy spectrum in the (very narrow) range between $\lambda_{\rm D}$ and $\rho_s$, and show that it is somewhat steeper than $-3$.
In the Appendix, we derive the disruption scale and the scalings for the energy spectrum in % to the case where $\beta_e \ll m_e/m_i$\footnote{This regime is applicable to the turbulence in the earth's aurora \citet{chaston2008}).}, and
the ``semicollisional" case, where the reconnection is enabled by resistivity, but the diffusion layer is much thinner than the ion scale $\rho_s$ -- a situation that is relevant to many laboratory experiments, e.g., TREX \citep{forest2015} and FLARE \citep{flare2014}, as well as in hybrid kinetic simulations \citep[e.g.,][]{parashar2009,kunz2014,cerri2017,cerri2017b}.
%The paper is organized as follows. In Section \ref{sec:turb}, we will review the turbulent scalings obtained in a simple model of Alfv\'enic turbulence \citep{ms16}. We will discuss collisionless reconnection in Section \ref{sec:tear}, before deriving the disruption scale $\lambda_{\rm D}$ in Section \ref{sec:scale}. Finally, in Section \ref{sec:disrange}, we will derive approximate scalings for the energy spectrum in the range between $\lambda_{\rm D}$ and $\rho_s$. 

\section{Alfv\'enic turbulence model}\label{sec:turb}
In the theory of intermittent Alfv\'enic turbulence of \cite{ms16}, the turbulence is modelled as an ensemble of structures, each of which is characterised by an Elsasser amplitude $\delta z$ and three characteristic scales: $\lpar$ (parallel), $\lambda$ (perpendicular) and $\xi$ (fluctuation-direction). We normalise these variables by their values at the outer scale:
\beq
\delta \hat{z} = \frac{\delta z}{\overline{\delta z}}, \quad \hlam = \frac{\lambda}{L_\perp}, \quad \hlpar=\frac{\lpar}{\Lpar}, \quad \hat{\xi} = \frac{\xi}{L_\perp},\label{eq:normalized}
\eeq
where $\overline{\delta z}$ is the outer-scale fluctuation amplitude, and $L_\perp$ and $L_\parallel$ are the perpendicular and parallel outer scales. In the following, we will treat $\hlam$ as a parameter (i.e. we are conditioning on $\hlam$): the distribution of $\delta \hat{z}$ depends on $\hlam$, and $\hat{\xi}$ and $\hlpar$ are calculated from $\delta\hat{z}$ and $\hlam$. It is assumed that the turbulence is critically balanced already at the outer scale. The normalised amplitude is given by 
\beq
\delta \hat{z} \sim \Lambda^q,\label{eq:pois}
\eeq
where $q$ is a Poisson-distributed random variable,
\beq
P(q) = \frac{\mu^q}{q!}e^{-\mu},\label{eq:pdf}
\eeq
with mean $\mu = -\ln\hlam$,\footnote{In the theory of \cite{ms16}, a slightly more complicated distribution is posited, but we ignore this nuance here and postulate (\ref{eq:pdf}).} and $\Lambda=1/\sqrt{2}$ is a dimensionless constant (which \citealt{ms16} called $\beta$, but which we here rename to avoid confusion with $\beta_e$). The scalings of  perpendicular structure functions are then given by
\beq
\langle \delta \hat{z}^n \rangle \sim \hlam^{\zeta_n^\perp},
\quad
\zeta_n^\perp = 1-\Lambda^n.\label{eq:zetan}
\eeq
The fluctuation-direction scale $\hat{\xi}$ is related to the amplitude via
\beq
\hat{\xi} \sim \hlam^{1/2} \Lambda^q,\label{eq:xiq}
\eeq
while the parallel scale depends only on $\hlam$:
\beq
\hlpar \sim \hlam^{1/2}.\label{eq:lpar}
\eeq
Following \cite{msc_disruption}, we define the ``effective amplitude" of structures that dominate the $n$-th order perpendicular structure function:
\beq
\delta \hat{z}[n]\equiv\langle \delta \hat{z}^n \rangle^{1/n} \sim \hlam^{\zeta_n^\perp/n}.\label{eq:effm}
\eeq
The effective amplitude $\delta \hat{z}[n]$ is a strictly increasing function of $n$, and so $n$ may be used as a convenient proxy for the amplitude of the structures at a given scale. The scalings for three interesting cases can be immediately obtained from (\ref{eq:effm}): first,
\beq
\delta \hat{z}[\infty] \sim 1
\eeq
describes the ``most intense" structures, whose amplitude is independent of scale; secondly,
\beq
\delta \hat{z}[2]\sim \hlam^{1/4}
\eeq
describes the fluctuations that dominate the second-order structure function and the energy spectrum, and thus determine the spectral index; finally, the ``bulk" fluctuations are described by $n\to0$, and their amplitudes scale as 
\beq
\delta \hat{z}[n\to0] \sim \hlam^{-\ln \Lambda}.
\eeq
We will also need an expression for the (effective) fluctuation-direction scale for the $n$-th order fluctuations, given by
\beq
\hat{\xi}[n] \sim \hlam^{1/2}\delta\hat{z}[n] \sim \hlam^{1/2+\zom},\label{eq:xi}
\eeq
and for the cascade time, 
\beq
\tau_{\rm C} \sim \frac{\xi}{\delta z} \sim \frac{L_\perp}{\overline{\delta z}} \hlam^{1/2}.\label{eq:tc}
\eeq
One can easily see from (\ref{eq:xi}) that the structures are anisotropic in the perpendicular plane, with $\xi\gg\lambda$, and that the higher-amplitude structures are more anisotropic, consistent with numerical evidence \citep{rcb,mallet3d}. 

We now have all the information needed about the turbulent structures to determine whether they can be disrupted by tearing. 
%\footnote{For $\delta u > \delta b$, the Kelvin-Helmholtz mode would disrupt the sheets much faster than the tearing mode. However, this situation does not occur, since the vortex-stretching terms for the different Elsasser fields $\vzp^\pm$ have opposite sign \citep{zhdankin16}, meaning that ``current sheets" are more common than ``shear layers" in RMHD turbulence, i.e. $\delta u < \delta b$, and the Kelvin-Helmholtz instability of the sheets is stabilized.}.

\section{Collisionless tearing mode}\label{sec:tear}
Scalings for the low-$\beta_e$ collisionless tearing mode are reviewed in Appendix B.3 of \cite{zocco2011}. Our sheet-like turbulent structures have a width $\lambda$ and a length $\xi$ in the perpendicular plane. We will assume that the perturbed magnetic field reverses across the structure $\delta b \sim \delta z$. There is also a velocity perturbation $\delta u$ associated with the $\delta z$. If the situation were that $\delta u > \delta b$, the Kelvin-Helmholtz instability would disrupt the sheets much faster than the tearing mode. However, this situation does not typically occur, because the vortex-stretching terms for the different Elsasser fields $\vzp^\pm$ have opposite sign \citep{zhdankin16intcy}, meaning that ``current sheets" are more common than ``shear layers" in RMHD turbulence, i.e., $\delta u < \delta b$ (this is also true in the solar wind; see, e.g., \citealt{chen2016,wicksalign}), and the Kelvin-Helmholtz instability is naturally stabilised \citep{chandrasekharbook}. For simplicity, we assume that the velocity fluctuations $\delta u \lesssim \delta b$ present in the sheet-like structures do not significantly affect the dynamics that we will describe in this paper.

The structure of the collisionless tearing mode involves three scales: the perpendicular scale $\lambda$ of the turbulent structure, and a nested inner layer, where two-fluid effects become important at the ion sound scale $\rho_s$, while flux unfreezing happens due to electron inertia in a thinner layer controlled by the electron inertial scale $d_e = c/\omega_{pe}\ll \rho_s$, where $\omega_{pe} = \sqrt{4\pi n_e e^2/m_e}$ is the electron plasma frequency. The tearing instability's growth rates will, therefore, involve all of these scales. The scalings that we use here will cease to apply if $\rho_s \lesssim d_e$ (i.e., $\beta_e \lesssim m_e/m_i$), at which point the ion scale becomes unimportant, and if $\beta_e\gtrsim 1$, when the flux unfreezing happens at the electron gyroradius $\rho_e = d_e\sqrt{\beta_e}$, rather than at $d_e$. This means that we are restricting ourselves to ``moderately" small beta, $1 \gg \beta_e \gg m_e/m_i$.
%The scalings we outline here are only valid when $\beta_e > m_e/m_i$, i.e., not so small that $d_e > \rho_s$, at which point the physics at ion scale $\rho_s$ is irrelevant to the tearing mode.

We will assume a Harris-sheet-like equilibrium \citep{harris1962}\footnote{In \citet{loureiroboldyrev_cless}, a more general class of equilibria is considered, which slightly affects the resulting scalings for the disruption scales and spectra.}. For long-wavelength modes ($k\ll 1/\lambda$), the instability parameter $\Delta'$ is given by 
\beq
\Delta' \lambda \approx \frac{1}{k\lambda}.\label{eq:longwave}
\eeq
For $\Delta'\delta_{\rm in} \ll 1$, where $\delta_{\rm in}$ is the width of the inner layer, the linear growth rate and the inner-layer width are
\beq
\gamma_> \sim k \delta z \frac{d_e \rho_s\Delta'}{\lambda}\sim {\delta z} \frac{d_e\rho_s}{\lambda^3}, \quad \delta_{\rm in} \sim d_e\rho_s^{1/2}\Delta'^{1/2}.\label{eq:trans}
\eeq
%where $d_e = c/\omega_{pe}$ is the electron inertial scale, and the ion sound scale $\rho_s=\rho_i\sqrt{Z T_e/2T_i}$. 
For $\Delta'\delta_{\rm in} \sim 1$, they are
\beq
\gamma_< \sim k\delta z \frac{d_e^{1/3} \rho_s^{2/3}}{\lambda},\quad \delta_{\rm in} \sim d_e^{2/3}\rho_s^{1/3}.\label{eq:coppi}
\eeq
The wavenumber $k_{\rm tr}$ of the transition between these two regimes can be found by balancing the two expressions for the growth rate, giving
\beq
k_{\rm tr} \sim \frac{d_e^{2/3}\rho_s^{1/3}}{\lambda^2}.
\eeq
For $k<k_{\rm tr}$, the growth rate is $\gamma_<$, while for $k>k_{\rm tr}$, it is $\gamma_>$, which is independent of wavenumber.
%. An important difference from the RMHD tearing mode\footnote{In the RMHD case, $\gamma_>\propto k^{-2/5}$ while $\gamma_< \propto k^{2/3}$, and so the maximum growth rate is attained at the RMHD equivalent of $k_{\rm tr}$.}  is that $\gamma_>$ is independent of wavenumber. 
This breaks down when $k\sim1/\lambda$, because (\ref{eq:longwave}) ceases to apply -- but, since $\xi\gg\lambda$, this only happens for a very large number of islands. Therefore, the maximum growth rate is attained for all $k>k_{\rm tr}$, and is given simply by $\gamma_>$. Thus there is always a mode with the maximum growth rate and a large enough wavenumber to fit into a sheet of any length $\xi>\lambda$. This is somewhat different from the resistive-RMHD case studied by \cite{msc_disruption}, in which $\gamma_>^{\rm RMHD}\propto k^{-2/5}$, while $\gamma^{\rm RMHD}_< \propto k^{2/3}$, and the maximum growth rate is attained at the transitional wavenumber.

The linear-growth stage of tearing ends when the width of the islands reaches $\delta_{\rm in}$, which decreases with increasing $k$ for $k>k_{\rm tr}$. Thus at the end of the linear stage, the largest islands are produced by the mode with $k\sim k_{\rm tr}$, and so, despite the independence of the linear growth rate on $k$, we can assume that this transitional mode dominates the nonlinear dynamics.
%, and the reconnection proceeds nonlinearly. The presence of kinetic Alfv\'en waves at ion scales $\sim \rho_s$ is known to make the reconnection fast (i.e. independent of microphysical scales) by opening up the effective width of the reconnecting region \citep{Shay1999,rogers2001}, so $\gamma_{\rm nl} \propto \delta z/\lambda$. This
% is
%, in fact observed numerically \citep{loureiro2013,numata2015}, and
% is always faster than the linear rate $\gamma_>$, since $\rho_s,d_e< \lambda$, and so the time $\tau_{\rm D}$ needed to disrupt the sheet-like structure by reconnection is given approximately by
We assume that the $X$-points between the islands then collapse quickly (i.e., on a timescale at most comparable to $\gamma_>^{-1}$), circularising the islands and forming a set of flux ropes of width $\lambda$, as appears to be consistent with numerical evidence \citep{loureiro2013}. \footnote{One can see that this is indeed what happens if, at the end of the nonlinear stage, the islands of width $\delta_{\rm in}$ and length $k_{\rm tr}^{-1}$ circularise at constant area: their width after circularization is $w_{\rm circ} \sim \sqrt{\delta_{\rm in} k_{\rm tr}^{-1}} \sim \lambda$.} Since these structures are as wide as the original sheet, the latter should at this point be disrupted and broken up -- being effectively replaced by a set of flux ropes.
The scale of these ropes parallel to the (exact) magnetic field is set as usual by critical balance. Since we assume that the $X$-point collapse is at least as fast as the linear tearing stage, we estimate the disruption time using the linear growth rate (\ref{eq:trans}) of the tearing mode:
\beq
\tau_{\rm D} \sim \gamma_>^{-1}.\label{eq:td}
\eeq

It is important to point out that the restriction to low $\beta_e$ limits the applicability of our conclusions in the solar wind, where, more often than not, $\beta_e \sim 1$, but our results will be more relevant to the turbulence closer to the Sun and in the corona: indeed, at the perihelion (approximately 10 solar radii) of the upcoming Parker Solar Probe mission, $\beta_e \approx 0.01$, at least in fast-solar-wind streams \citep{chandran2011}. Moreover, the growth-rate scaling (\ref{eq:trans}) appears to be quite robust even at moderately large $\beta_e$: \cite{numata2015} showed that, keeping all other parameters fixed, $\gamma_>\propto \beta_e^{-1/2}\propto d_e$ up to at least $\beta_e = 10$, in agreement with (\ref{eq:trans}), and despite the width of the reconnecting layer being set by $\rho_e$ rather than $d_e$. Therefore, we expect our conclusions to be at least qualitatively relevant at $\beta_e\sim 1$.

\section{Disruption scale}\label{sec:scale}
A sheet-like structure will be disrupted if its nonlinear cascade time (\ref{eq:tc}) is longer than its disruption time (\ref{eq:td}). The disruption scale $\hld$ is then determined by demanding
\beq
\frac{\tau_{\rm C}}{\tau_{\rm D}} \sim {\xi}\frac{d_e \rho_s}{\lambda^3}\gtrsim1.\label{eq:tctdcomp}
\eeq
Using (\ref{eq:xi}) and (\ref{eq:effm}), we find that, for $n$-th order aligned structures, this inequality is satisfied for
\beq
\hlam \lesssim \hld[n] \sim \left(\frac{d_e \rho_s}{L_\perp^2}\right)^{\frac{2}{5}\left(1-2\zeta_n^\perp/5n\right)^{-1}}.\label{eq:hld}
\eeq
The scale $\hld$ is an increasing function of $n$. It is largest for the most intense structures, with $n\to\infty$, for which
\beq
\hld[\infty] \sim  \left(\frac{d_e \rho_s}{L_\perp^2}\right)^{{2}/{5}}.\label{eq:hldinf}
\eeq
The scale at which the $n=2$ structures, which determine the scaling of the second-order structure function and the energy spectrum, are disrupted is\footnote{This scaling has also been independently derived by \cite{loureiroboldyrev_cless}.}
\beq
\hld[2] \sim  \left(\frac{d_e \rho_s}{L_\perp^2}\right)^{{4}/{9}},\label{eq:m2}
\eeq
and, finally, the bulk fluctuations ($n\to0$) are disrupted at
\beq
\hld[0] \sim \left(\frac{d_e \rho_s}{L_\perp^2}\right)^{0.46}.\label{eq:m0}
\eeq
The disruption may effectively be thought of as taking place over a narrow %\footnote{For realistic values of $d_e$ and $\rho_s$ corresponding to the solar wind, the range between $\hld[0]$ and $\hld[\infty]$ is somewhat less than half a decade.} 
range of scales between $\hld[\infty]$ and $\hld[0]$, with $\hld[2]$ as a good representative.
The disruption will only be relevant if any of these scales is larger than the scale at which the waves become dispersive, i.e., 
\begin{align}
\frac{\lamd[n]}{\rho_s} &\sim \left(\frac{d_e}{\rho_s}\right)^{\frac{2}{5}\left(1-2\zeta_n^\perp/5n\right)^{-1}}\left(\frac{L_\perp}{\rho_s}\right)^{1-\frac{4}{5}\left(1-2\zeta_n^\perp/5n\right)^{-1}}\nonumber\\
&\sim \left(\frac{m_i}{m_e}\frac{\beta_e}{Z}\right)^{-\frac{1}{5}\left(1-2\zeta_n^\perp/5n\right)^{-1}}\left(\frac{L_\perp}{\rho_s}\right)^{1-\frac{4}{5}\left(1-2\zeta_n^\perp/5n\right)^{-1}}\gtrsim1.
\end{align}
This gives us a critical $\beta_e$ for structures at any given $n$ to be disrupted:
\beq
\beta_e \lesssim \beta_e^{\rm crit}[n] \sim Z \frac{m_e}{m_i} \left(\frac{L_\perp}{\rho_s}\right)^{1-2\zeta^\perp_n/n}.\label{eq:bcrit}
\eeq
For the $n=2$ structures,
\beq
\beta_e^{\rm crit}[2] \sim Z \frac{m_e}{m_i} \left(\frac{L_\perp}{\rho_s}\right)^{1/2},\label{eq:bcrit2}
\eeq
while for the most intense fluctuations ($n\to\infty$),
\beq
\beta_e^{\rm crit}[\infty] \sim Z \frac{m_e}{m_i} \frac{L_\perp}{\rho_s}.\label{eq:bcritinf}
\eeq
 It is interesting to note that despite the fact that the dependence of $\hld[n]$ on $n$  does not appear to be very strong [the exponents in (\ref{eq:hldinf}), (\ref{eq:m2}) and (\ref{eq:m0}) are close together, at $0.40$, $0.44$, and $0.46$ respectively], the dependence of $\beta_e^{\rm crit}$ on $L_\perp/\rho_s$ is a strong function of $n$. 
 
In the solar wind, typically $L_\perp/\rho_s \approx 10^3$ \citep{chen2016}, and so $\beta_e^{\rm crit}[2] \sim 10^{-2}$, which is rare at 1AU but should be rather common closer to the Sun in the region to be explored by the Parker Solar Probe \citep{spp2016,chandran2011}. On the other hand, $\beta_e^{\rm crit}[\infty] \sim 1$, and so one might expect the most intense sheet-like structures to become unstable to the onset of reconnection even at moderate $\beta_e$. Flux-rope-like ``Alfv\'en vortex" structures extended in the parallel direction were indeed observed at ion scales in the solar wind by \cite{perrone2016} and \cite{lion2016}. It is tempting to identify the structures produced by the disruption due to tearing with these observations\footnote{\citet{cerri2017} and \citet{franci2017} observed reconnection onset and the formation of chains of multiple islands in their hybrid simulations of 2D kinetic turbulence. It is tempting to identify the island chains in their simulations with the structures that we predict here, but it should be noted that their simulations are 2D, and do not model electron inertia (or Ohmic resistivity), so the reconnection mechanism is quite different, and only a qualitative comparison can be made.}. We will study the structures produced by the disruption process in the next section, quantifying the relationship between their amplitude and scale, and further examining their importance as a function of $\beta_e$ and of $\rho_s/L_\perp$.

Finally, let us set aside for a moment the precise values of $\beta_e^{\rm crit}$ for which we predict that disruption happens, and focus instead on the scaling of the break in the energy spectrum, $\hld[2]$, with physical parameters, i.e., the dependence of $\hld[2]$ on $\beta$. \citet{chenbreak2014} observed that at low $\beta_i$, the break scale of solar-wind turbulence appeared to scale as $d_i \propto \rho_i/\sqrt{\beta_i}$, in contradiction with expectations based on the linear physics of low-$\beta$ plasmas \citep{schektome2009}. Here we predict (ignoring the factor of $(L_\perp/\rho_s)^{1/9}$, which barely changes with the relevant physical parameters)
\beq
\frac{\lamd[2]}{\rho_s} \propto \beta_e^{-2/9} \quad \Rightarrow \quad \frac{\lamd[2]}{d_i} \propto \left(\beta_i\frac{T_e}{T_i}\right)^{5/18}.
\eeq
For disruption due to reconnection to explain the anomalous break scale observed by \citet{chenbreak2014}, there would therefore have to be correlations between $\beta_i$ and $T_e/T_i$ in their chosen intervals (namely, $T_i/T_e \propto \beta_i$ to match precisely). Encouragingly, in their data it does appear that the lower-$\beta_i$ intervals are associated with markedly higher $T_e/T_i$.

 %\footnote{
%It should be pointed out that the growth rate scalings Eqs.~(\ref{eq:trans}) and (\ref{eq:coppi}) were derived, as we have mentioned, for $\beta_e \ll 1$ \citep{zocco2011}. However, the growth rate scalings appear to be quite robust%, even at moderately large $\beta_e$
%: \cite{numata2015} showed that, keeping all other parameters fixed, $\gamma_>\propto \beta_e^{-1/2}\propto d_e$ up to at least $\beta_e = 10$, in agreement with Eq.~(\ref{eq:trans}); despite the width of the reconnecting layer being set by $\rho_e$ rather than $d_e$.%}
\section{Statistical properties of flux ropes}\label{sec:stat}
The dependence (\ref{eq:hld}) of $\hld$ on $n$ is one way of quantifying the scales at which the disrupted structures appear. In this section, we recast our calculation, treating the amplitude of the fluctuation as a random variable, i.e., we return to (\ref{eq:pois}), and determine what fraction of the aligned structures remain undisrupted at any given scale, in terms of $q$ (we remind the reader that this is an integer distributed as a Poisson random variable with mean $\langle q\rangle=\mu=-\ln\hlam$). For a structure to be disrupted, we again demand (\ref{eq:tctdcomp})
and, using (\ref{eq:xiq}),
 find that %(\ref{eq:tctdstat}) becomes
\beq
\hlam^{-5/2}\Lambda^q\frac{d_e \rho_s}{L_\perp^2}\gtrsim1.
\eeq
This is satisfied for
\begin{align}
q\lesssim q_{\rm D} &= \frac{({5}/{2})\ln\hlam - \ln\left({d_e\rho_s}/{L_\perp^2}\right)}{\ln\Lambda}\nonumber\\&= \frac{({5}/{2})\ln\hlam - 2\ln\left({\rho_s}/{L_\perp}\right)+({1}/{2})\ln\left({\beta_e m_i}/{Z}{m_e}\right)}{\ln\Lambda}.\label{eq:qd}
\end{align}
\subsection{Filling factor of aligned turbulence}\label{sec:ff}
At any given scale, the filling factor of sheet-like, aligned structures that have not been affected by the disruption process (i.e., the probability of encountering them) is given by
\beq
f_0(\hlam) = P(q>q_D) = 1-\sum_{q=0}^{\lfloor\qd\rfloor}P(q),\label{eq:fp}
\eeq
where the distribution of $q$ is given by (\ref{eq:pdf}).
Similarly, the disruption causes a fractional reduction of energy contained in aligned sheet-like structures that is, using (\ref{eq:zetan}) and (\ref{eq:pois}),
\begin{align}
f_2(\hlam) = 1- \frac{1}{\langle\delta \hat{z}^2\rangle}\sum_{q=0}^{\lfloor\qd\rfloor} \delta \hat{z}^2P(q)=1- \frac{1}{\hlam^{1/2}}{\sum_{q=0}^{\lfloor\qd\rfloor}2^{-q}P(q)}.\label{eq:fe}
\end{align}
Obviously, (\ref{eq:fp}) and (\ref{eq:fe}) can only be considered quantitatively good estimates if $f_0$ and $f_2$ are close to unity, i.e., if the overall ``RMHD ensemble" (described in Section \ref{sec:turb}) is not significantly altered. 

\begin{figure}
\begin{center}
\includegraphics[width=0.7\linewidth]{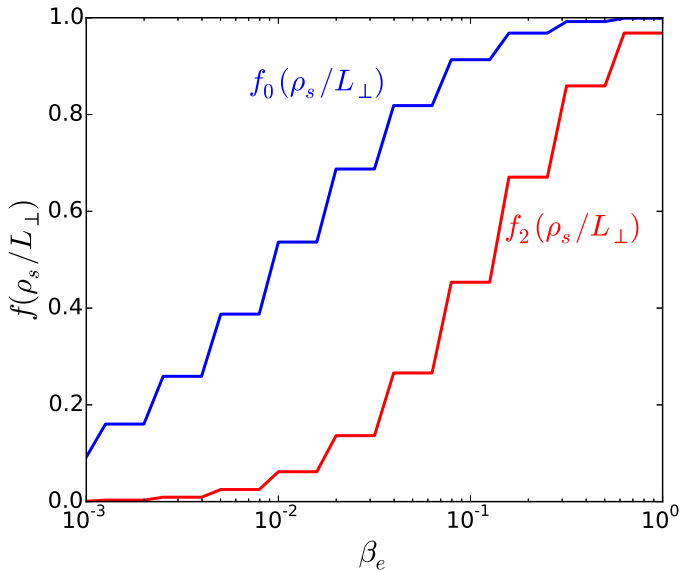}
\end{center}
\caption{The filling factor of the aligned (i.e., undisrupted) structures at the ion sound scale $f_0(\rho_s/L_\perp)$ (blue) and fraction $f_2(\rho_s/L_\perp)$ of energy in them (red), plotted as a function of $\beta_e$. We have taken $\rho_s/L_\perp = 10^{-3}$, a reasonable value for the solar wind. Since $q$ is an integer but $q_{\rm D}$ is not, the sums (\ref{eq:fp}) and (\ref{eq:fe}) are performed up to $\lfloor q_{\rm D}\rfloor$, resulting in the discontinuities shown in the plot. In reality, of course, $f_0$ and $f_2$ will be smooth. %\textbf{The region $\beta_e\geq1$ is shaded because the asymptotic tearing-mode growth rates quoted in Section \ref{sec:tear} are not rigorously valid there.}
\label{fig:fpfe}}
\end{figure}

Using (\ref{eq:qd}), both $f_0$ and $f_2$ may be calculated numerically, as functions of $\hlam$, $\rho_s/L_\perp$, and $\beta_e$. A particularly interesting case is $\lambda=\rho_s$, since this quantifies the cumulative effect of reconnection on the turbulence at the ion scale. Figure \ref{fig:fpfe} shows the dependence of $f_0(\rho_s/L_\perp)$ and $f_2(\rho_s/L_\perp)$ on $\beta_e$. The effect of disruption becomes more important at smaller $\beta_e$. Note that the amount of energy in the undisrupted structures at the ion scale is significantly reduced for values of $\beta_e$ somewhat larger than $\beta_e^{\rm crit}[2]$ given by (\ref{eq:bcrit2}). This suggests that, in practice, the turbulence is significantly affected by the disruption at only moderately small $\beta_e$: e.g., for $\beta_e \sim 0.1$, only around half of the energy that would be in sheets without disruption actually makes it to the ion scale, despite only around $10\%$ by volume of the turbulence being disrupted. 

\subsection{Amplitude of flux ropes}\label{sec:amps}

We have proposed that the sheet-like structures with $q\sim\qd$ disrupted by tearing at the scale $\lambda$ are converted into circular flux ropes, with perpendicular scale $\lambda$. We will assume that, just after they are created, they have the same amplitude as the sheet-like structure that produced them:
\beq
\delta \hat{z}_{\rm fr} \sim \Lambda^\qd.\label{eq:frqd}
\eeq
The flux ropes will not stay around for long: they will interact with each other and the remaining sheet-like structures, cascade, align, and form smaller, more sheet-like structures: this process will be studied in the next section. However, we can predict the relationship between the amplitude $\delta z_{\rm fr}$ and radius $\lambda$ of the \emph{newly created} flux ropes: upon inserting (\ref{eq:qd}) into (\ref{eq:frqd}), we get
\beq
\delta \hat{z}_{\rm fr} \sim \hlam^{5/2}\left(\frac{L_\perp}{\rho_s}\right)^{2}\left(\frac{\beta_e}{Z}\frac{m_i}{m_e}\right)^{1/2}.\label{eq:dzfr}
\eeq
It is important to note that this is not a prediction for the scaling of any structure function or the spectrum of the disrupted turbulence (which will be worked out in the next section); rather, this is a relationship describing \emph{individual} flux ropes upon their formation within an aligned structure.

Thus, provided that $\beta_e$ and $\rho_s/L_\perp$ are small enough that $\qd>0$ before the cascade reaches $\rho_s$, i.e., $\beta_e<\beta_e^{\rm crit}[\infty]$ given by (\ref{eq:bcritinf}), there should be a strong relationship between the amplitude and radius of the structures at scales between $\hld[\infty]$, given by (\ref{eq:hldinf}), and $\rho_s$: indeed, the scaling (\ref{eq:dzfr}) is very steep with $\lambda$, and the flux ropes with the largest radius $\lamd[\infty]$ also have the largest amplitude, $\delta z_{\rm fr} \sim \overline{\delta z}$. The scaling (\ref{eq:dzfr}) could in principle be tested against observations such as those reported by \citet{perrone2016}, who observed 12 ``Alfv\'en vortices" with diameters between $5\rho_i$ and $17\rho_i$ (as part of a sample of over 100 coherent structures of different types).

\section{Disruption-range turbulence}\label{sec:disrange}
In a realistic situation relevant to coronal or solar-wind turbulence, the separation between $\lamd$ and $\rho_s$ is so small that it would be challenging to establish a robust distinction between these two scales. It is nonetheless interesting to speculate on the nature of the turbulence in the interval $\lamd \gg \lambda \gg \rho_s$ in the asymptotic case where $L_\perp\to\infty$.

As described in Section \ref{sec:tear}, we expect the disruption process to convert the sheet-like structures just above $\lamd$ into flux-rope-like structures just below $\lamd$. These are roughly circular in the perpendicular plane, with radius $\lamd$, but extended in the parallel direction due to critical balance. 
In order to treat this ``disruption range" properly, we would need to account for the intermittency of both sheets and flux ropes. We do not attempt such a treatment in this paper. Instead, we develop a simpler model, in which we take the sheets and flux ropes to be effectively volume filling, and the sheets to have the properties of the $n=2$ structures of Section~\ref{sec:turb} (since we would like to explore the scaling of the energy spectrum below $\lamd$). For simplicity of notation, we will drop the ``$[n=2]$" argument of all relevant quantities. 
%We will completely ignore intermittency, and adopt scalings based solely on the $n=2$ fluctuations, since we would like to explore the scaling of the energy spectrum below $\lamd$ --- we therefore drop the ``$[n=2]$" argument of all relevant quantities. 
A ``characteristic fluctuation amplitude" $\delta z_{1,-}$ for the turbulence just below $\lamd$ may be defined by assuming that there is negligible dissipation during the disruption process, so that the energy flux stays constant across the disruption scale\footnote{
The same assumption is made in the treatment of recursive disruptions in Section \ref{sec:recurse}. This assumption may appear questionable (although, for a collisionless plasma, not necessarily impossible, since the reconnection itself is mediated by electron inertia and does not require dissipation), especially if the reconnection process were to proceed literally all the way to saturation, reconnecting all of the available flux and generating vigorous outflows, which can be Landau damped \citep{loureiro2013,tenbarge2013,navarro2016}. Since the nonlinear cascade time, the linear tearing time, and the time for the islands to grow to the same width as the aligned structure that spawned them and thus disrupt it, are all of the same order, how much dissipation is likely to happen before this disruption may be a quantitative issue contingent on the precise, order-unity relationships between these times (note that \citealt{loureiro2013} report peak dissipation nearly 10 Alfv\'en times after peak reconnection). \citet{bl2017} and \citet{loureiroboldyrev_cless} resolve this by assuming that the tearing mode can disrupt its mother sheet without needing to produce a perturbation of the magnetic field comparable in size to the fields associated with the sheet --- and thus without dissipating much energy. We do not see how, dynamically, this can happen, since the process of disruption is presumably the very same nonlinear process that leads to islands perturbing the sheet finitely. Two further observations regarding dissipation in reconnection events are that (i) it is not necessarily the case that dissipation of energy caused during the disruption of a sheet of scale $\lambda$ can be viewed as happening {\it at} scale $\lambda$, rather than as being part of the overall energy transfer towards smaller scales, where the dissipation actually occurs; (ii) how effective Landau damping is in dissipating energy in a truly collisionless,  turbulent plasma is an open question \citep{schekochihin2016}.}. 
Further assuming (radically) that the turbulence just below $\lamd$ loses all its dynamic alignment, the scale $\xi$ is no longer relevant and the only perpendicular scale in the problem is $\lamd$:
\beq
\epsilon \sim \frac{\overline{\delta z}^3}{L_\perp} \sim \frac{\delta z_{1,-}^3}{\lambda_{\rm D}},\label{eq:constflux1}
\eeq
giving
\beq
\delta \hat{z}_{1,-} \sim \hld^{1/3}.\label{eq:minus}
\eeq
Note that this is smaller than the amplitude of the aligned turbulence at the same scale:
\beq
\delta \hat{z}_{1,+} \sim \hld^{1/4}.\label{eq:plus}
\eeq
The definition (\ref{eq:minus}) might appear to be in contradiction with our earlier assumption (\ref{eq:frqd}) that the flux ropes should be formed with the same amplitude as their ``mother sheet". Indeed, if one took (\ref{eq:frqd}), (\ref{eq:minus}) and (\ref{eq:plus}) together, they imply that the flux ropes created by a disrupting sheet fill only a fraction of the mother sheet's volume, contradicting our supposition above. However, (\ref{eq:minus}) is not meant to be the amplitude of any individual structure, but rather an effective estimate that would on average give (\ref{eq:constflux1}). In this sense, there is no difference between (\ref{eq:minus}) or (\ref{eq:plus}) and the usual ``twiddle" relations in Kolmogorov-style turbulence phenomenologies relying on constant energy flux and ignoring intermittency and local imbalance: the amplitude (\ref{eq:minus}) is an estimate that effectively absorbs within itself the filling fraction (probability of occurence) of energetic structures that contribute to averages, as well as the (unknown) details of how precisely the nonlinear interactions within or between flux ropes (and between flux ropes and ambient turbulence) actually occur. In the case of flux ropes, it is clear that to make a connection between (\ref{eq:frqd}) and any average quantity, one would need to take into account the fact that they are clearly less volume-filling than their mother sheets, have shorter lifetimes, and might mainly cascade due to interactions between different types of structures. We leapfrog these issues with the aid of the requirement (\ref{eq:constflux1}) that energy flux should stay the same. This will allow us to make progress and develop a simple model in this Section: (\ref{eq:minus}) will define the ``outer scale amplitude" for the Alfv\'enic cascade below $\lamd$. Intermittency is known to be of crucial importance in Alfv\'enic turbulence \citep{Chandran14,ms16}, and so a more rigorous model incorporating intermittency should be the subject of future work.

Following \cite{msc_disruption}, the turbulence just below $\lamd$ should behave just like the usual Alfv\'enic turbulence described in Section \ref{sec:turb} (since $\lamd \gg \rho_s$): the flux ropes will interact with each other and the rest of the turbulence, causing a cascade to smaller scales\footnote{
The nonlinear evolution of the flux ropes is likely to be more complex than simply pairs of them merging into a single larger flux rope \citep{fermo2010}. First, once the sheet is disrupted, the islands are not forced to stay at the location of the original sheet and so are not constrained to interact in a quasi-1D setting, moving along the sheet \citep[cf.][]{uzdensky2010,loureiro2012}, or, indeed, to interact only with each other, rather than with the ambient turbulence. Secondly, since all this happens in three dimensions, they can cross, shear each other, or break up in more ways than are available to 2D plasmoids in a 1D sheet.
}. In the course of this secondary cascade, the turbulence will again start to dynamically align and form sheet-like structures, and may eventually be disrupted by the onset of reconnection at a secondary disruption scale $\lambda_{\rm{D},2}$, at which the whole process repeats --- provided that $\lambda_{\rm{D},2}>\rho_s$. Therefore, the turbulence between the first disruption scale $\lamd$ (which we will now rechristen $\lambda_{\rm{D},1}$) and $\rho_s$ is characterised by a sequence of disruptions, between which the turbulence re-aligns. We will now show that this recursive disruption process is unlikely to be fully realised, and usually terminates after only one disruption.
\subsection{Recursive disruption?}\label{sec:recurse}
The sequence of disruptions described above may be understood in terms of a recursion relation. After the $(i-1)$st disruption, the turbulence undergoes the $i$th ``mini-cascade", with ``outer-scale" values of the relevant quantities given by the values just below the $(i-1)$st disruption scale $\lambda_{\rm{D},i-1}$:
\beq
\overline{\delta z} \to \delta z_{i-1,-}, \quad L_\perp \to \lambda_{{\rm D},i-1}.\label{eq:subs}
\eeq
Substituting this into (\ref{eq:m2}) and normalising by $L_\perp$, we obtain the $i$-th disruption scale
\beq
\hldp\sim \hldpm^{1/9}\left(\frac{d_e\rho_s}{L_\perp^2}\right)^{4/9}
\sim \left(\frac{d_e\rho_s}{L_\perp^2}\right)^{\frac{1}{2}\left(1-\frac{1}{9^i}\right)}.\label{eq:hldp}
%\sum_{j=0}^{i-1}\left(\frac{1}{9}\right)^{j}}.
\eeq
Unlike in the resistive case \citep{msc_disruption}, this sequence will always terminate after a finite number of disruptions, because $\rho_s>d_e$ and so eventually $\lambda_{{\rm D},i}<\rho_s$. The number of disruptions is given by the greatest $i = i_{\rm max}$ for which
\beq
\frac{\lambda_{\rm{D},i}}{\rho_s} \sim \left(\frac{d_e}{\rho_s}\right)^{\frac{1}{2}\left(1-\frac{1}{9^i}\right)}%{\frac{4}{9}\sum_{j=0}^{i-1}\left(\frac{1}{9}\right)^{j}}
\left(\frac{L_\perp}{\rho_s}\right)^{\frac{1}{9^i}}>1.\label{eq:ndis}
%{1-\frac{8}{9}\sum_{j=0}^{i-1}\left(\frac{1}{9}\right)^{j}} > 1.
\eeq
It is obvious from the exponent of $L_\perp/\rho_s$ in (\ref{eq:ndis}) that, for there to be more than one disruption, $L_\perp/\rho_s$ must be unrealistically large. Namely, (\ref{eq:ndis}) may be solved for $i_{\rm max}$, showing an extremely weak dependence on $L_\perp/\rho_s$:
\beq
i_{\rm max} = \frac{\ln\left[1+\frac{2\ln(L_\perp/\rho_s)}{\ln(\rho_s/d_e)}\right]}{2\ln3}.
\eeq
%As $d_e\to\rho_s$, the number of disruptions grows to infinity. This describes the transition to the ``ultra-low beta" regime where $\beta_e < m_e/m_i$,
With $d_e/\rho_s$ kept constant as $L_\perp/\rho_s\to\infty$, the number of disruptions grows extremely slowly, so, in a moderately-low-$\beta_e$ situation where $d_e / \rho_s < 1$, it is very unlikely that more than one disruption will occur. 

\subsection{Effective spectral index below $\lamd$: many disruptions}\label{sec:many}

Nevertheless, in the spirit of asymptotic fantasising, let us determine the effective spectral index in a scale range featuring many disruptions. With the substitutions (\ref{eq:subs}), the characteristic fluctuation amplitude just below each disruption scale is [cf. (\ref{eq:minus})]
\beq
\delta \hat{z}_{i,-} \sim \hldp^{1/3}.\label{eq:generalminus}
\eeq
Therefore, the fluctuation amplitude associated with the $i$th aligning ``mini-cascade" between disruption scales $\hldpm$ and $\hldp$ is 
\beq
\delta \hat{z}_{i} \sim \delta \hat{z}_{-,{i-1}}\left(\frac{\hlam}{\hldpm}\right)^{1/4}\sim \hldpm^{1/3}\left(\frac{\hlam}{\hldpm}\right)^{1/4}.\label{eq:alin}
\eeq
The characteristic aspect ratio of the turbulence (equivalently, the inverse alignment angle) between disruptions is then
\beq
\frac{\xi_i}{\lambda} \sim \left(\frac{\hlam}{\hldpm}\right)^{-1/4},\label{eq:aspn}
\eeq
which is much smaller than the value ($\hlam^{-1/4}$) that would have been attained without the disruptions. The ``coarse-grained" fluctuation amplitude calculated at the scale just above the $i$th disruption is [cf. (\ref{eq:plus})]
\beq
\delta \hat{z}_{i,+} \sim \hldpm^{1/3} \left(\frac{\hldp}{\hldpm}\right)^{1/4}\sim \hldp \left(\frac{d_e\rho_s}{L_\perp^2}\right)^{-1/3},\label{eq:upper}
\eeq
where we have used (\ref{eq:hldp}) to obtain the second expression. Since (\ref{eq:upper}) is larger than (\ref{eq:generalminus}), our model spectrum looks like a sloping staircase, with (\ref{eq:upper}) providing an upper envelope for the true scaling of the fluctuation amplitude (the true spectrum and structure function will not, of course, have discontinuities). Thus, $\delta z_{+,i} \propto \lambda_{{\rm D}, i}$, and so the effective scaling exponent of the fluctuation amplitudes is $\alpha_{\rm eff} = 1$. This implies a spectral index of $-1-2\alpha_{\rm eff} = -3$, slightly steeper than the observed spectral index $\approx -2.8$ below the ion scales in strong kinetic-Alfv\'en-wave turbulence in the solar wind \citep{alexandrova2009,chenkaw2010,sahraoui2010}.

Exactly the same scaling is (of course) found by observing that $\tc \sim \gamma_>^{-1}$ on the coarse-grained points $\hldp$: then, constancy of energy flux through all scales implies that (cf. \citealt{bl2017})
\beq
\epsilon \sim \frac{\delta z_{i,+}^{2}}{\tau_{\rm{C}}} \sim \gamma_{>} \delta z_{i,+}^2 \propto \frac{\delta z_{i,+}^3}{\lambda_{{\rm D},i}^3}\sim \text{const.}\quad\Rightarrow\quad \delta z_{+,i} \propto \lambda_{{\rm D}, i}.\label{eq:constflux}
\eeq

\subsection{Effective spectral index below $\lamd$: one disruption}\label{sec:onedis}
For realistic values of $L_\perp$ and $\rho_s$, there is only one disruption, i.e., $i_{\rm max} = 1$. The aligning cascade below $\lamd$ then gives an amplitude at $\rho_s$ of
\beq
\delta \hat{z}_{\rho_s} \sim \hld^{1/3}\left(\frac{\rho_s}{\lamd}\right)^{1/4},
\eeq
using (\ref{eq:alin}). The characteristic aspect ratio of the turbulence at $\rho_s$ is, using (\ref{eq:aspn}),
\beq
\frac{\xi_1}{\rho_s} \sim \left(\frac{\rho_s}{\lamd}\right)^{-1/4} \sim \left(\frac{m_i}{m_e}\frac{\beta_e}{Z}\right)^{-1/18}\left(\frac{\rho_s}{L_\perp}\right)^{-1/36},\label{eq:asp1}
\eeq
which is approximately unity for any realistic set of parameters. Therefore, the turbulence at the ion scale, i.e., at the largest scales in the kinetic Alfv\'en wave cascade, will be very different depending on the presence or absence of the disruption: namely, it will either be nearly isotropic (in the perpendicular plane), as per (\ref{eq:aspn}), or highly anisotropic (aligned) with aspect ratio $(\rho_s/L_\perp)^{-1/4}$, respectively. In reality, there may be a mixture of both types of structures, as is suggested by the discussion in Section \ref{sec:ff} --- so the reduction in the alignment may not be as drastic as suggested by the extreme estimate (\ref{eq:asp1}).

The effective scaling of the fluctuation amplitudes between $\lamd$ and $\rho_s$ will be steeper than $\alpha_{\rm eff} = 1$ derived in Section \ref{sec:many}, because the recursive disruptions are cut off by the presence of the ion scale $\rho_s$. The effective scaling in this case is given by
\beq
\alpha_{\rm eff} = \frac{\log(\delta \hat{z}_{1,+}/\delta \hat{z}_{\rho_s})}{\log(\lamd/\rho_s)}.\label{eq:aeffreal}
\eeq
This ranges between $1 \leq \alpha_{\rm eff} < \infty$, increasing the closer $\rho_s$ gets to $\lamd$. %In reality, it cannot be steeper than the characteristic scaling exponent of the individual flux-rope like structures themselves, which are thought to have an energy spectrum of approximately $k_{\perp}^{-4}$ \citep{Alexandrova2008}.
%attains its lowest value of $\alpha_{\rm eff, min} \approx 1$ at the point where $\rho_s = \lambda_{\rm{D},2}+\delta$, as $\delta\to0$, and its highest value $\alpha_{\rm eff,max}\to\infty$, when $\rho_s = \lambda_{\rm{D},1}-\delta$, as $\delta\to0$. 
Thus the effective spectral index in a ``realistic" short range of scales between $\lamd$ and $\rho_s$, with only one disruption, may be somewhat steeper than $-3$, and may depend on $\beta_e$ (i.e., it is not universal). Here again, the caveat that disruption does not in fact occur at a single scale or in every aligned structure implies that the spectrum should not be as dramatically steepened as (\ref{eq:aeffreal}) suggests. A spectrum slightly steeper than $-3$ is probably a reasonable expectation.

Short intervals of steep spectra are indeed sometimes observed near the ion scales in the solar wind: \citet{sahraoui2010} and \citet{lion2016} report a spectral index close to $-4$ in a small ``transition range" of scales near the ion gyroradius. \cite{lion2016} attribute this spectral index to Alfv\'en vortices \citep{Alexandrova2008} with scales a few times the ion gyroradius.% which, as we discussed in Section \ref{sec:stat}, may be similar to the flux-ropes that are the saturated state of the disruption process outlined here. It is tempting to speculate that the vortex structures observed by \citet{lion2016} are the flux-ropes produced by the disruption mechanism outlined here, and to this end, it would be very interesting to study systematically the presence or absence of the transition range and of the vortices/flux-ropes (and their statistical properties -- see Section \ref{sec:stat}) as a function of $\beta_e$ and $\rho_s/L_\perp$. If the transition range is really caused by the disruption process described here, we expect it to appear at somewhat low $\beta_e$.

\section{Discussion}
Models of strong Alfv\'enic turbulence that incorporate dynamic alignment predict that the turbulent structures become progressively more sheet-like at smaller scales \citep{boldyrev,Chandran14,ms16}. This suggests that at some sufficiently small scale $\lamd$, the cascade time of the structures may be slower than the time required to disrupt them via magnetic reconnection. For resistive RMHD, this scale was calculated by \cite{msc_disruption} and \cite{loureiroboldyrev}. In this paper, we have extended this idea to the case of a weakly collisional, low-$\beta_e$ plasma, in which the reconnection is due to electron inertia, rather than resistivity, and two-fluid effects become important at ion scales. We find that there is again a critical scale, $\hld\sim L_\perp^{1/9}(d_e\rho_s)^{4/9}$, below which the sheet-like structures are destroyed by reconnection. For sufficiently low electron beta, and sufficiently large scale separation between the outer scale $L_\perp$ and the ion sound scale $\rho_s$, this scale lies in the inertial range: $\lamd>\rho_s$. The break in the energy spectrum of turbulence in a low-$\beta$ collisionless plasma can thus occur at a larger scale than expected based on linear physics of wave modes --- it does indeed do so in the solar wind \citep{bourouaine2012,chenbreak2014}, although the observed scaling of the break scale with $\beta_i$ appears to be stronger than we are able to predict here, unless there is some systematic correlation of the electron-ion temperature ratio with $\beta_i$. %This situation, where the reconnection process depends on two dimensionless parameters ($\beta_e$ and $L_\perp/\rho_s$) is very different than in the resistive RMHD case \citep{msc_disruption}, where the disruption scale is always larger than the resistive cutoff, because they both depend on only one dimensionless parameter, the magnetic Reynolds number $\SL= \overline{\delta z}L_\perp/\eta$.
 %both the disruption scale and the resistive cutoff depend only on the magnetic Reynolds number $\SL = \overline{\delta z}L_\perp/\eta$. 
 
We have argued that between $\lamd$ and $\rho_s$, the spectral index of the turbulent fluctuations should be steeper than $-3$. A steep ``transition range" around the ion scale is indeed sometimes observed in the real solar wind \citep{sahraoui2010}. This has been attributed to the presence of Alfv\'en vortices \citep{Alexandrova2008,lion2016}. These may be similar to the flux-rope-like structures that we envision in this paper to be the product of the disruption of the aligned cascade, and so disruption via tearing may be a physical reason for the presence of Alfv\'en vortices at ion scales in the solar wind. We predict that such structures should become very unlikely above a certain $\beta_e$ [given by (\ref{eq:bcritinf})], and that as $\beta_e$ decreases, the proportion of the volume at the ion scale filled with aligned, undisrupted structures decreases, as does the amount of energy contained in them (Section \ref{sec:ff}). We also propose the relationship (\ref{eq:dzfr}) between the amplitude and radius of the individual flux-rope structures. This could potentially be tested by solar-wind observations of the kind performed by \citet{perrone2016}. 

For the Alfv\'enic turbulence to be disrupted by reconnection, we need $\lamd > \rho_s$. This inequality translates into a requirement that the electron plasma beta must be less than some critical value $\beta_e^{\rm crit}$, given by (\ref{eq:bcrit}), which depends on the ratio $L_\perp/\rho_s$ of the outer scale to the ion scale and on the amplitude of the structures being considered. In the solar wind, $L_\perp/\rho_s \approx 10^3$, and we find that for fluctuations of moderate amplitude (ones that dominate the energy spectrum), $\beta_e^{\rm crit} \sim 0.01$, while for the most intense (but rare and intermittent) fluctuations, $\beta_e^{\rm crit}\sim 1$. Thus, we expect only the most intense structures to be disrupted in the solar wind at 1AU, where $\beta_e \sim 1$. Closer to the sun, $\beta_e$ may be lower \citep{chandran2011}, and the disruption process becomes more effective. 

The turbulence at the ion scale is significantly different depending on whether disruption due to the onset of reconnection can occur or not. Above $\beta_e^{\rm crit}$, sheet-like Alfv\'enic structures with a large aspect ratio will reach the ion scale without disruption. Below $\beta_e^{\rm crit}$, the disruption should occur, and turbulence at the ion scale should become much less anisotropic (less aligned) in the perpendicular plane (see Section \ref{sec:onedis}). Thus, the nature of the turbulence at the ion scale, which provides the starting point for the sub-ion-scale kinetic-Alfv\'en-wave turbulence, depends crucially on whether the disruption process occurs.

%There are many extensions to the model proposed here that could be made. The most obvious one is to study the case when $\lamd < \rho_s$: for this, we would need to know something about the nature of the structures formed by the turbulence at scales $k_\perp\rho_s\gtrsim1$. There are some indications that these structures are also sheet-like \citep{boldyrevkaw2012,wan2016}, and if the dependence of the fluctuation amplitude and aspect ratio of these sheets on scale $\lambda$ was known, one could extend this model quite easily. %What we have shown here is that for sufficiently low $\beta_e$, the disruption via reconnection does in fact happen at a scale $\lamd>\rho_s$, and the sheet-like structures naturally formed by dynamically aligning Alfv\'enic turbulence are destroyed by magnetic reconnection. This causes a steepening of the spectrum between $\lamd$ and the ion scale $\rho_s$, and dramatically alters the morphology of the structures that provide the starting point for the sub-$\rho_i$ kinetic Alfv\'en wave turbulence.

\section*{Acknowledgements} We thank N. Loureiro and O. Alexandrova for useful conversations. The work of A.M. was supported by NSF grant AGS-1624501. B.D.G.C. was supported by NASA grants NNX15AI80G, NNX16AG81G, and NNX17AI18G, and NSF grant PHY-1500041. The work of A.A.S. was supported in part by grants from UK  STFC and EPSRC.
\appendix
%\section{The KREHM equations}\label{app:krehm}
\section{Semicollisional disruption} \label{app:other}
Here, we will replicate some of the calculations done in the main text, but this time for a low-$\beta$ ``semicollisional", large-guide-field regime where the width of the diffusive layer $\delta_{\rm in}$ is smaller than $\rho_s$, but is controlled by resistivity rather than by electron inertia. For the turbulence, this means that the ion scale is greater than the resistive scale that would have provided the dissipative cutoff in fully collisional MHD,
\beq
\frac{\rho_s}{L_\perp} \gg \SL^{-3/4},
\eeq
but the electron-ion collision rate is nevertheless much larger than the nonlinear cascade rate of the fluctuations, or the growth rate of the tearing mode,
\beq
\nu_{ei}\gg  \tc^{-1}\sim \gamma.
\eeq
This means that the the flux conservation is broken by Ohmic resistivity $\eta = \nu_{ei} d_e^2$ rather than by electron inertia, but the two-fluid effects are still important. This situation is relevant to many laboratory plasmas, for example TREX \citep{forest2015} and FLARE \citep{flare2014}, as well as in hybrid-kinetic simulations that do not model the electron inertia \citep[e.g.,][]{parashar2009,kunz2014,cerri2017,cerri2017b}.

\subsection{Semicollisional tearing mode}
Growth rates of the tearing mode in this regime are reviewed in Appendix B.5 of \citet{zocco2011}. We will assume that the turbulent structures are still given by our RMHD turbulence model summarized in Section \ref{sec:turb}; viz., they have a width $\lambda$ and length $\xi$ in the perpendicular plane, and a perturbed-magnetic-field reversal $\delta b \sim \delta z$ across $\lambda$. The tearing mode in this regime, like its collisionless cousin in Section \ref{sec:tear}, involves three-scale physics: the (unstable) ``equilibrium" at the scale of the turbulent structure $\lambda$, two-fluid effects at around the ion sound scale $\rho_s$, and flux unfreezing in an inner layer of width $\delta_{\rm in}\ll\rho_s$, controlled by resistivity. 

We will again consider the long-wavelength limit, $k\lambda \ll 1$, and assume (\ref{eq:longwave}).
For $\Delta'\delta_{\rm{in}} \ll 1$, the linear growth rate and the inner layer's width are
\beq
\gamma_> \sim k\delta z \left(\Delta'\rho_s\right)^{2/3}\left(k\lambda S_\lambda\right)^{-1/3} \sim \frac{\delta z}{\lambda} \left(\frac{\rho_s}{\lambda}\right)^{2/3} S_\lambda^{-1/3},\quad\delta_{\rm in} \sim \lambda(\Delta'\rho_s)^{1/6}(k\lambda S_\lambda)^{-1/3},\label{eq:scgp}
\eeq
where $S_\lambda \doteq \delta z \lambda / \eta$ is the Lundquist number based on scale $\lambda$. Note that, as in the collisionless case, $\gamma_>$ is independent of $k$. For $\Delta'\delta_{\rm in}\sim 1$,
\beq
\gamma_< \sim k\delta z \left(\frac{\rho_s}{\lambda}\right)^{4/7} \left(k\lambda S_\lambda\right)^{-1/7}, \quad \delta_{\rm in} \sim \lambda\left(\frac{\rho_s}{\lambda}\right)^{1/7}(k \lambda S_\lambda)^{-2/7}.
\eeq
The transitional wavenumber $k_{\rm tr}$ between these two regimes may be found by balancing the two expressions for the growth rate, giving
\beq
k_{\rm tr}\lambda \sim \left(\frac{\rho_s}{\lambda}\right)^{1/9}S_\lambda^{-2/9}.
\eeq
For all $k> k_{\rm tr}$, the growth rate is given by $\gamma_>$, so there is always a mode with the maximum growth rate that has a short enough wavelength to fit into the sheet. 

The linear stage of tearing ends when the width of the islands reaches $\delta_{\rm in}$. This is largest for $k<k_{\rm tr}$, so, at the end of the linear stage, the largest islands are again produced by the mode with $k\sim k_{\rm tr}$, similarly to the collisionless case. We will, therefore, again assume that this mode dominates the nonlinear dynamics. We will also again assume that the $X$-points between the islands collapse on a timescale at least as short as the linear stage, and that the islands circularise forming a set of flux ropes. Similarly to the collisionless case, if they do so at constant area, their width is
\beq
w_{\rm circ} \sim \sqrt{\delta_{\rm in} k_{\rm tr}^{-1}} \sim \lambda.
\eeq
Since we are assuming that the circularisation is at least as fast as the linear stage, we can again estimate the disruption time using the linear growth rate (\ref{eq:scgp}),
\beq
\tau_{\rm D} \sim \gamma_>^{-1}.
\eeq
%We again assume that the reconnection is fast, since this is due to two-fluid effects at the ion scale $\rho_s$, retained in this regime, and so after the linear growth ends, the reconnection quickly saturates the mode, forming circularised islands of scale $\lambda$. Thus, $\tau_{\rm D} \sim \gamma_>^{-1}$.
\subsection{Disruption scale}
Disruption of an aligned structure will occur if
\beq
\frac{\tau_{\rm C}}{\tau_{\rm D}} \sim \frac{\xi}{\lambda} \left(\frac{\rho_s}{\lambda}\right)^{2/3} S_\lambda^{-1/3} \gtrsim 1.
\eeq
Therefore, the aligned structures are disrupted for
\beq
\hlam < \hld \sim \left[\left(\frac{\rho_s}{L_\perp}\right)^2\SL^{-1}\right]^{\frac{2}{9}\left(1-4\zeta_\perp^n/9n\right)^{-1}}.
\eeq
%where $\SL \doteq \overline{\delta z}{L_\perp}/\eta$ is the outer scale Lundquist number or magnetic Reynolds number. 
For the $n=2$ structures, this gives
\beq
\hld[2] \sim  \left(\frac{\rho_s}{L_\perp}\right)^{1/2}\SL^{-1/4},\label{eq:schld2}
\eeq
while for the most intense structures ($n=\infty$),
\beq
\hld[\infty] \sim \left(\frac{\rho_s}{L_\perp}\right)^{4/9}\SL^{-2/9}.
\eeq
The disruption scale is larger than $\rho_s$ when $\SL$ is below an $n$-dependent critical value:
\beq
\SL < \SL^{\rm crit} \sim \left(\frac{L_\perp}{\rho_s}\right)^{\frac{5}{2}\left(1-4\zeta_n/5n\right)}.
\eeq
For the $n=2$ structures,
\beq
\SL^{\rm crit}[2] \sim \left(\frac{L_\perp}{\rho_s}\right)^2,
\eeq
and for the $n=\infty$ structures,
\beq
\SL^{\rm crit}[\infty] \sim \left(\frac{L_\perp}{\rho_s}\right)^{5/2}.
\eeq
Thus, the disruption becomes progressively more important to the aligned Alfv\'enic turbulence above the ion scale as $\SL$ decreases --- until $\SL^{-3/4} \gtrsim \rho_s/L_\perp$, at which point the semicollisional tearing mode scalings are no longer valid and the fully resistive regime studied in \citet{msc_disruption} is reached.

Note that, from (\ref{eq:schld2}),
\beq
\frac{\lamd[2]}{\rho_s} \sim \left(\frac{\rho_s}{L_\perp}\right)^{-1/2}\SL^{-1/4}.
\eeq
This can be compared with attainable values of $\SL$ and $\rho_s/L_\perp$ in laboratory experiments: for example, according to \citet{forest2015}, the TREX experiment is able to access $10^3 \lesssim \SL\lesssim10^5$ and $1\lesssim L_\perp/\rho_s \lesssim10^2$. This results in values of the disruption scale in the interval
\beq
2 \gtrsim \frac{\lamd[2]}{\rho_s} \gtrsim 0.1,
\eeq
and so it is at least plausible that the disruption of Alfv\'enic turbulence by semicollisional tearing could be observed in such an experiment.

\subsection{Amplitude of flux ropes}
Instead of characterising the structures by $n$, we can, similarly to what we did with the collisionless case in Section \ref{sec:stat}, return to (\ref{eq:pois}) and treat the amplitude of a fluctuation as a random variable. Via an analogous derivation, for disruption to occur, we must have
\beq
\frac{\tc}{\tau_{\rm D}} \sim \frac{\xi}{\lambda} \left(\frac{\rho_s}{\lambda}\right)^{2/3} S_\lambda^{-1/3} \gtrsim 1\quad\Rightarrow\quad\hlam^{-3/2} \Lambda^{2q/3} \left(\frac{\rho_s}{L_\perp}\right)^{2/3}\SL^{-1/3}\gtrsim1,
\eeq
which is satisfied for
\begin{align}
q\lesssim \qd = \frac{\ln\hlam - (4/9)\ln(\rho_s/L_\perp) + (2/9)\ln\SL}{\ln\Lambda}.\label{eq:qdsc}
\end{align}
One may use this and (\ref{eq:pdf}) to calculate the semicollisional versions of the filling fraction of aligned structures, $f_0$, and the remaining fraction of their energy, $f_2$ [Section \ref{sec:ff}, (\ref{eq:fp}) and (\ref{eq:fe}), respectively], and show that as $\SL$ decreases, these both become smaller. 

If we again assume that the newly created flux ropes have the same amplitude as their mother sheets [see (\ref{eq:frqd})], then, using (\ref{eq:qdsc}), we obtain a relationship between $\delta z_{\rm fr}$ and $\lambda$ of the flux ropes just after they are created:
\beq
\delta \hat{z}_{\rm fr} \sim \hlam \left(\frac{\rho_s}{L_\perp}\right)^{-4/9} \SL^{2/9}.
\eeq
This relationship between the radii and amplitudes of individual flux ropes produced by the disruption of turbulent structures could in principle be tested in laboratory plasma devices, or in numerical simulations. Note that, as in Section \ref{sec:amps}, the largest flux ropes are produced with radius $\hld[n=\infty]$ and have amplitude $\delta z_{\rm fr} \sim \overline{\delta z}$. 
\subsection{Recursive disruption?}
Similarly to the collisionless case (Section \ref{sec:disrange}), the flux-rope-like fluctuations just below the disruption scale should seed a new Alfv\'enic cascade to smaller scales, aligning, and potentially disrupting again, and so on recursively until the ion scale $\rho_s$ is reached. Let us again focus on $n=2$, i.e., on the fluctuations that determine the scaling of the energy spectrum. Again, (\ref{eq:minus}) and (\ref{eq:subs}) describe the $i$th ``mini-cascade". The recursive relation between successive disruption scales is, analogously to (\ref{eq:hldp}):
\beq
\hldp\sim \hldpm^{1/6}\left(\frac{\rho_s}{L_\perp}\right)^{1/2}\SL^{-1/4} \sim\left(\frac{\rho_s}{L_\perp}\SL^{-1/2}\right)^{\frac{3}{5}\left(1-\frac{1}{6^i}\right)}.\label{eq:schldp}
\eeq
This is valid for all $i$ for which $\hldp > \rho_s/L_\perp$, giving a condition for the $i$th disruption to be realised:
\beq
\frac{\lambda_{{\rm D},i}}{\rho_s}\sim \left(\SL^{3/4}\frac{\rho_s}{L_\perp}\right)^{-\frac{2}{5}\left(1-\frac{1}{6^i}\right)}\left(\frac{L_\perp}{\rho_s}\right)^{\frac{1}{6^i}}\gtrsim1.\label{eq:sc_ldrhos}
\eeq
This inequality is always violated for
\beq
i_{\rm max} \sim \frac{\ln\left(1+\frac{5\ln(L_\perp/\rho_s)}{2\ln(\SL^{3/4}\rho_s/{L_\perp})}\right)}{\ln6}<\infty,
\eeq
unless $\rho_s/L_\perp < \SL^{-3/4}$, at which point the semicollisional scalings are no longer valid and the system is in
the fully resistive RMHD regime studied in \citet{msc_disruption}. We again see that for there to be more than one disruption above $\rho_s$, $L_\perp/\rho_s$ must be unrealistically large. Thus, the semicollisional regime, like the collisionless regime, is always characterised by a limited number of disruptions -- usually a single disruption, for realistic parameters.

\subsection{Effective spectral index below $\lamd$}
Suppose, despite the arguments in the previous section as to the real-world irrelevance of this situation, that the range of scales between $\lamd$ and $\rho_s$ is asymptotically broad, and there are a large number of disruptions, $i_{\rm max} \gg 1$ (and also the semicollisional regime remains operative, i.e., $\rho_s/L_\perp > \SL^{-3/4}$). Using (\ref{eq:subs}), the turbulent amplitude just below each disruption is again (\ref{eq:generalminus}). Between successive disruption scales, the aligning cascade causes the turbulent amplitude to follow (\ref{eq:alin}), and the aspect ratio of structures is given by (\ref{eq:aspn}), but now with the disruption scales $\hldp$ given by (\ref{eq:schldp}). Again, the aspect ratio of the structures is greatly reduced by the disruption. The ``coarse-grained" amplitudes calculated at the scales just above the $i$th disruption is [cf. (\ref{eq:alin})],
\beq
\delta \hat{z}_{i,+} \sim \hldpm^{1/3}\left(\frac{\hldp}{\hldpm}\right)^{1/4}
\sim \hldp^{3/4} \left(\frac{\rho_s}{L_\perp}\right)^{1/4}\SL^{-1/8}.\label{eq:scalings_sc}
\eeq
Thus, in the semicollisional case, the effective scaling exponent of the fluctuation amplitude is $\alpha_{\rm eff} = 3/4$, implying a spectral index of $-2.5$. Exactly the same scaling is obtained by noticing that on the coarse-grained points $\hldp$, the cascade time is just set by the linear growth rate of the tearing mode (\ref{eq:scgp}), $\tc \sim \gamma_>^{-1}$ and requiring that $\epsilon \sim \delta z^2/\tc\sim \text{const.}$ [cf. (\ref{eq:constflux})].

For realistic values of $\rho_s/L_\perp$ and $\SL$, there will be at most a single disruption before $\rho_s$. Essentially an identical argument to the one given in Section \ref{sec:onedis} implies that, due to the cutoff at $\rho_s$, the scaling of the fluctuation amplitudes may be nonuniversal, somewhat steeper than (\ref{eq:scalings_sc}), and depend on the values of the physical parameters. As in the collisionless case, the presence of the disruption causes the structures at the ion scale to be much less anisotropic within the perpendicular plane than they would have been had reconnection not interfered.

%As in the collisionless case, the PRESENCE of the disruption causes the structures at the ion scale to be MUCH LESS ANISOTROPIC THAN THEY WOULD HAVE BEEN HAD RECONNECTION NOT INTERFERED. 
%At each disruption, $\gamma_>\tau_{\rm C} \sim 1$, and so on the coarse-grained set of points $\hldp$, using the constancy of the energy flux through scale,
%\beq
%\epsilon \sim \frac{\delta z_{i,+}^2}{\tau_{\rm C}} \sim \gamma_>\delta z_{i,+}^3 \propto \frac{\delta z_{i,+}^{8/3}}{\lambda_{{\rm D},i}^2},
%\eeq
%we find $\delta z_{i,+}\propto \lambda_{{\rm D},i}^{3/4}$. Thus, in the semicollisional regime $\alpha_{{\rm eff}} = 3/4$ and the spectrum between $\lamd$ and $\rho_s$ should be approximately $E(k_\perp) \propto k_\perp^{-2.5}$. As we pointed out in the collisionless case (see Section \ref{sec:onedis}), the true spectrum may be somewhat steeper than this since in most situations there will be only one disruption before the ion scale $\rho_s$ is reached.
%\subsection{Other regimes}

\bibliographystyle{jpp}

\bibliography{mainbib2}

\end{document}